\documentclass{svproc}
\usepackage{url,graphicx,bm,amsmath,amsfonts,multicol,algorithm, algorithmic}

\begin{document}
\mainmatter
\title{Improved algorithm for neuronal ensemble inference by Monte Carlo method}
\titlerunning{Improved algorithm for neuronal ensemble inference}
\author{Shun Kimura \and Koujin Takeda}
\tocauthor{Shun Kimura, Koujin Takeda}
\institute{Department of Mechanical Systems Engineering, \\
           Graduate school of Science and Engineering, \\
           Ibaraki University\\
\email{Correspondence:koujin.takeda.kt@vc.ibaraki.ac.jp}
}

\maketitle

\begin{abstract}
Neuronal ensemble inference is one of the significant problems in the study of biological neural networks. Various methods have been proposed for ensemble inference from their activity data taken experimentally. Here we focus on Bayesian inference approach for ensembles with generative model, which was proposed in recent work. However, this method requires large computational cost, and the result sometimes gets stuck in bad local maximum solution of Bayesian inference. In this work, we give improved Bayesian inference algorithm for these problems. We modify ensemble generation rule in Markov chain Monte Carlo method, and introduce the idea of simulated annealing for hyperparameter control. We also compare the performance of ensemble inference between our algorithm and the original one.

\end{abstract}
\section{Introduction}
In recent study of biological neural network, advanced recording technologies such as calcium imaging or functional magnetic resonance imaging enable us to obtain neuronal activity data from thousands of neurons simultaneously. Such activity data will reveal features of neural network. For instance, neurons in the same neuronal ensemble tend to fire synchronously\cite{Pascal}\cite{Vitor:Sergio:Miguel:Sidarta}, therefore identification of ensembles is significant for understanding whole neural network structure and its dynamical behavior. In fact, there are some studies on neural network structure using ensemble information\cite{Rainer:Christopher:Gilles}\cite{J.matias:Simo:Shrikanth:Satu}.

Several statistical methods are known for ensemble identification in activity data. For instance, principal component analysis or singular value decomposition can identify ensembles\cite{Sebastian:Thomas:Veronica:Adrien:Mathieu:German}. Their advantage is that they can effectively reduce dimension and volume of activity data, which are very large in general. On the other hand, these methods require prior knowledge such as the number of ensembles. There is an alternative approach using graph theory, where neuronal activity is expressed as nodes in graph. From such graph, ensemble activity can be extracted by graph clustering method such as spectral clustering\cite{Hiromu:Takayuki:Sayuri:Takeshi:Yuishi}, while this method is basically applied to static data and neglects dynamical behavior.

One of the strategies to overcome above-mentioned problems is Bayesian modeling for neuronal activity. In recent work, generative model of ensemble activity was proposed by Bayesian inference framework\cite{Giovanni:Thomas:Martin}. Using Markov chain Monte Carlo (MCMC) method, this enables us to infer neuronal ensembles and their dynamical behavior. However, this requires large computational cost in MCMC, and the result sometimes gets stuck in bad local maximum solution of Bayesian inference.

In this work, for reduction of computational cost and bad local maximum problem, we propose an improved algorithm for neuronal ensemble identification. First, we change the update rule in MCMC for controlling the number of ensembles. Second, we introduce the idea of simulated annealing for hyperparameter control. We also compare our algorithm and the original one in terms of ensemble identification using synthetic data, and discuss the advantage of our method.

\section{Theory}

\subsection{Bayesian inference model}
Here we outline the framework of Bayesian inference. We basically follow the notation in the original paper\cite{Giovanni:Thomas:Martin}. See it for the detail. Each neuron has the label $i \in \{ 1, 2, \ldots N\}$, and $N$ is the total number of neurons. The variable $k \in \{1, 2, \ldots M\}$ is the time step, and $M$ is the size of time frame. The variable $\mu \in \{1, 2, \ldots A\}$ is the label of neuronal ensemble, and $A$ is the total number of ensembles. The $i$th neuron has neuronal membership label to an ensemble, $t_i \in \{1, 2, \ldots, A\}$, and activity variable $s_{ik} \in \{0, 1\}$ at time $k$. The $\mu$th ensemble has ensemble activity variable $\omega_{k \mu} \in \{0, 1\}$ at time $k$. The state $1$ for binary variable means active (firing) neuron or neuronal ensemble, and the state $0$ is inactive. 

With these variables, we give generative model for neuronal activity as the conditional joint probability, 
\begin{eqnarray}
P( \bm t, \bm \omega, \bm s|\bm n, \bm p, \bm \lambda) 
 & \propto & \left(
             \prod_{i=1}^{N} n_{t_{i}}
             \right)
             \cdot
             \left(
             \prod_{\mu=1}^{A} \prod_{k=1}^{M} 
             p_{\mu}^{\omega_{k \mu}}
             (1-p_{\mu})^{1-\omega_{k \mu}} 
             \right)
             \nonumber \\
         & & \cdot 
             \left(
             \prod_{i=1}^{N} \prod_{k=1}^{M} 
             [\lambda_{t_{i}}(\omega_{kt_{i}})]^{s_{ik}} 
             [1-\lambda_{t_{i}}(\omega_{kt_{i}})]^{(1-s_{ik})}
             \right),
\end{eqnarray}
where boldface letter represents the set of variables (e.g. $\bm t = \{t_1,t_2, \ldots t_N\}$). The $\mu$th ensemble has activity rate $p_\mu$, which means activity of ensemble. It also has assign probability $n_{\mu}$, which describes how many neurons belong to this ensemble. Conditional activity rate $\lambda_{t_{i}}$ of the $i$th neuron for given ensemble activity $\omega_{kt_{i}}$ is defined as
\begin{eqnarray}
\lambda_{t_i}( \omega_{k t_i} ) = P(s_{ik}=1\ |\ \omega_{k t_i} )\ \ {\rm for}\ \ \omega_{k t_i} \in \{ 0,1 \}.
\label{condactivity}
\end{eqnarray}
Accordingly, inactivity rate is expressed as $1 - \lambda_{t_i}( \omega_{k t_i} ) = P(s_{ik}=0\ |\ \omega_{k t_i} )$. The parameter $\bm \lambda$ describes coherence or incoherence (=noise) between neuronal activity and ensemble activity. We assume that the priors of ensemble activity rate $\bm p$ and conditional activity rate $\bm \lambda$ are beta distribution (denoted by Beta), while the prior of assign probability $\bm n$ is Dirichlet distribution (by Dir), namely
\begin{eqnarray}
P (p_{\mu}) & = & {\rm Beta} \left( 
                             \alpha^{(p)}_{\mu}, \beta^{(p)}_{\mu} 
                             \right), \\
P (\lambda_{\mu}(z)) & = & {\rm Beta} \left( 
                                      \alpha_{z, \mu}^{(\lambda)},  
                                      \beta_{z, \mu}^{(\lambda)} 
                                      \right), \\
P(  n_{1}, \cdots, n_{A} ) & = & {\rm Dir} \left( 
                                           \alpha_{1}^{(n)},\cdots, 
                                           \alpha_{A}^{(n)} 
                                           \right),
\end{eqnarray}
where $\alpha^{(p)}_{\mu}, \beta^{(p)}_{\mu}, \alpha_{z,\mu}^{(\lambda)}, \beta_{z,\mu}^{(\lambda)}, \alpha_{\mu}^{(n)} $ ($z \in \{0,1\}, \mu \in \{1,2,\ldots, A\}$) are hyperparameters. The relation among variables/parameters in this generative model is represented graphically in Fig. \ref{Fig.graphical}A.

\begin{figure}[t]
\begin{picture}(0,250)
	\put(0,0){\includegraphics[scale=0.42]{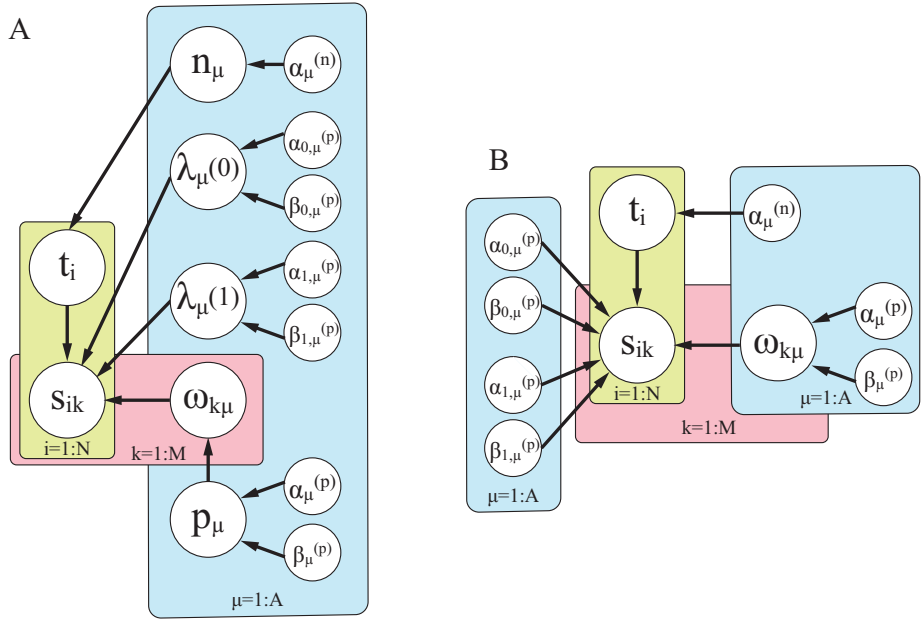}}
	\label{Fig.graphical}
\end{picture}
\caption{Graphical representation of the relation among variables/parameters in Bayesian inference model:
	    (A) the full model 
	    (B) the model after integrating out of
            $\bm n, \bm p, \bm \lambda$.}	              \label{Fig.graphical}
\end{figure}

For improvement of inference accuracy, we analytically integrate out the set of model parameters $\{ \bm n, \bm p, \bm \lambda \}$. Integration over these parameters leads to the joint probability as
\begin{eqnarray}
P(\bm t, \bm \omega, \bm s) 
& = &       \int d \bm n d \bm p d \bm \lambda\ 
            P(\bm t, \bm \omega, \bm s | \bm n, \bm p, \bm \lambda )
            P( \bm n, \bm p, \bm \lambda ) \nonumber \\
& \propto & \int d \bm n d \bm p d \bm \lambda
            \left(
            \prod_{i=1}^{N} n_{t_{i}}
            \right)
            \cdot
            {\rm Dir} \left( 
                      \alpha_{1}^{(n)}, \ldots, \alpha_{A}^{(n)}
			          \right)
		    \nonumber \\           
& &         \cdot 
            \left(
            \prod_{\mu=1}^{A} \prod_{k=1}^{M} 
            p_{\mu}^{\omega_{k\mu}}
            (1-p_{\mu})^{1-\omega_{k\mu}} 
            \right)
            \cdot 
            \left( 
            \prod_{\mu=1}^{A} {\rm Beta} 
            \left( 
            \alpha^{(p)}_{\mu}, \beta^{(p)}_{\mu}
			\right) 
			\right)
            \nonumber \\
& &         \cdot 
            \left( 
            \prod_{i=1}^{N} \prod_{k=1}^{M} 
            [\lambda_{t_{i}}(\omega_{k t_{i}})]^{s_{ik}} 
            [1-\lambda_{t_{i}}(\omega_{kt_{i}})]^{(1-s_{ik})}
            \right) \nonumber\\
& &         \cdot 
            \left( 
            \prod_{\mu=1}^{A} \prod_{z \in \{ 0,1 \}}
            {\rm Beta} \left( 
                       \alpha_{z, \mu}^{(\lambda)},
			           \beta_{z, \mu}^{(\lambda)}
			           \right) 
	        \right)	\nonumber \\		                
& = &       \left(
            \frac{\mathcal{B}(\alpha_1^{(n)}+G_1,\ 
                  \alpha_2^{(n)}+G_2, \ldots,\ \alpha_A^{(n)}+G_A)}
                 {\mathcal{B}(\alpha_1^{(n)}, \alpha_2^{(n)}, 
                  \ldots, \alpha_A^{(n)})}
            \right) \nonumber \\ 
& &         \cdot \prod_{\mu=1}^{A}
            \left\{
            \frac{B(H_{\mu}, \bar{H}_{\mu})}
                 {B(\alpha_{\mu}^{(p)}, \beta_{\mu}^{(p)})}
            \prod_{z \in \{ 0, 1 \}}
            \frac{B(T_{\mu}^{z1}, T_{\mu}^{z0})}
                 {B(\alpha_{z,\mu}^{(\lambda)}, \beta_{z,\mu}^{(\lambda)})}     
            \right\} .
\label{P(t, omega, s)}
\end{eqnarray}
Note that $B(\cdot, \cdot)$ is beta function, and $\mathcal{B}$ is defined by gamma functions as
\begin{equation}
\mathcal{B} ( x_{1}, \cdots, x_{A} )
\equiv \frac{\prod_{k=1}^A {\Gamma}(x_{k})}
            {{\Gamma}(\sum_{k=1}^A x_{k})}.
\end{equation}
In addition, we introduce several variables, 
\begin{eqnarray}
G_{\mu} & = & \sum_{i=1}^{N} \delta_{\mu, t_{i}},
\quad
H_{\mu} = \alpha^{(p)}_{\mu} + \sum_{k=1}^{M}\omega_{k\mu},
\quad
\bar{H}_{\mu} = \beta^{(p)}_{\mu} + \sum_{k=1}^{M}(1-\omega_{k\mu}), \nonumber \\
T_{\mu}^{z 1}
 &=& \alpha_{z, \mu}^{(\lambda)} + 
 \sum_{k=1}^{M} \left( \sum_{i \in \bm \mu} \delta_{z, \omega_{k\mu}} \delta_{1, s_{ik}} 
 \right),\ \
 T_{\mu}^{z 0}
 = \beta_{z, \mu}^{(\lambda)} + 
 \sum_{k=1}^{M} \left( \sum_{i \in \bm \mu} \delta_{z, \omega_{k\mu}} \delta_{0, s_{ik}} 
 \right), \nonumber \\
\end{eqnarray}
where $\delta_{ij}$ is Kronecker delta function, boldface $\bm \mu$ is the set of neurons in the $\mu$th ensemble, and $z \in \{0,1\}$. All variables are defined for the $\mu$th ensemble, and their meanings are as follows:
$G_\mu$ counts the number of neurons in the ensemble. $H_{\mu}$ and $\bar{H}_{\mu}$ indicate frequencies of active and inactive states, respectively. $T_{\mu}^{z1}$ and $T_{\mu}^{z0}$ measure coherence between ensemble activity and neuronal activity under the same superscript numbers, and incoherence (=noise) under different superscript numbers. The relation among variables/parameters in the model after integrating out of parameters $\{ \bm n, \bm p, \bm \lambda \}$ is represented in Fig. \ref{Fig.graphical}B.

The posterior $P(\bm t, \bm \omega| \bm s)$ can be constructed from this model. With this posterior, we can infer membership $\bm t$ and ensemble activity $\bm \omega$ from neuronal activity variable $\bm s$, which is obtained experimentally.

\subsection{Improvement of inference algorithm for the number of neuronal ensembles}

With the scheme mentioned above, we can obtain neuronal ensembles and ensembles activity by Bayesian inference. However, the number of possible neuronal states in all neurons is huge, therefore direct Bayesian inference is infeasible. 

For computational cost problem, we employ MCMC to evaluate maximum of posterior distribution. In the previous work\cite{Giovanni:Thomas:Martin} collapsed Gibbs sampling is used, where the number of ensembles $A$ is unknown and we also need to evaluate it. For inference of $A$, Dirichlet process (DP)\cite{Radford} is introduced, and we can vary $A$ till convergence of DP. However, we need to start with {\it large} $A$ as initial condition in DP. If we start with small $A$, we will get stuck at the solution of very few ensembles or without ensemble structure, which is supposed to be bad local maximum solution of Bayesian inference. Hence, this method still requires large computational cost at early stage of MCMC for successful inference, because the cost is proportional to $A$ at a given MCMC stage. (See Algorithm \ref{alg.update}.)

To cope with this problem, we propose a novel method. The differences from the previous work are summarized as follows:
\begin{itemize}
\item{
In our method, when new ensemble is created for increasing $A$ in DP, 
{\it multiple} neurons can move to new ensemble {\it simultaneously}, while 
single neuron can move to new ensemble in the original. Introduction of such simultaneous move will make
 new ensemble hard to vanish.  
}
\item{
We apply the idea of {\it simulated annealing} to transient probability to new ensemble in DP.}
\end{itemize}
As shown later, our method enables us to infer appropriate ensemble structure {\it without starting large $A$}. \\

We give the detail of our algorithm in the following. In our MCMC, we first update ensemble activity $\bm \omega$, then update ensemble membership label $\bm t$ by collapsed Gibbs sampling. This process is the same as the original. Next, we employ DP in order to increase/decrease the number of ensembles. Suppose that we have $A$ ensembles in the intermediate stage of MCMC. Destination ensemble of the $i$th neuron after MCMC update, denoted by $t_{i}^*$, is drawn from the probability,
\begin{equation}
q_{i}(t_{i}^*) = \begin{cases} \displaystyle
                \frac{G_{t_{i}^*}^{(\backslash i)}}
                     {q_{\alpha}^{[\gamma]} + N - 1}
                & {\rm for}\ \ t_{i}^* = 1, \ldots, A, \vspace{2mm} \\
                \displaystyle
                \frac{q_{\alpha}^{[\gamma]}} 
                     {q_{\alpha}^{[\gamma]} + N - 1} 
                & {\rm for}\ \ t_{i}^* = A + 1.
                \end{cases}
\label{DP}     
\end{equation}
The backslash $\backslash$ denotes the removal of a specific element, and $G_{t_{i}^*}^{(\backslash i)}$ means the number of neurons in the $t_{i}^*$th ensemble, where the $i$th neuron is not counted. Note that $\sum_{\mu=1}^A G_{\mu}^{(\backslash i)} = N-1$ and $q_{i}(t_{i}^*)$ satisfies $\sum_{\mu=1}^{A+1} q_{i}(t_{i}^*) = 1$. The parameter $q_{\alpha}^{[\gamma]}$ is proportional to transient probability to the new $(A+1)$th ensemble. As mentioned before, we apply the idea of simulated annealing to DP. In our method, the transient parameter at the $\gamma$th MCMC stage $q_{\alpha}^{[\gamma]}$ decays exponentially as 
\begin{equation}
q_{\alpha}^{[\gamma]} =  q_{\alpha}^{[0]}
                     e^{- \frac{\gamma}{\tau}},
\label{qdecay}
\end{equation}
where $\tau$ is decay constant. The idea of equation (\ref{qdecay}) is that the number of ensembles $A$ is changed frequently at early MCMC stage for exploring appropriate $A$, while the change is suppressed at late stage for convergence. If new ensemble is accepted in DP, we must generate new ensemble activity $\bm \omega$ and hyperparameters of new ensemble. We give hyperparameters of the new $(A+1)$th ensemble as arithmetic average of already-existing hyperparameters, because new hyperparameter should have the same scale as others for appropriate convergence of MCMC.
\begin{eqnarray}
\alpha_{A+1}^{(p)} & = & \frac{1}{A} \sum_{\mu=1}^{A} \alpha_{\mu}^{(p)}, \ \ 
\beta_{A+1}^{(p)} = \frac{1}{A} \sum_{\mu=1}^{A} \beta_{\mu}^{(p)}, \nonumber \\ 
\alpha_{z,A+1}^{(\lambda)} 
 & = & \frac{1}{A} \sum_{\mu=1}^{A} \alpha_{z,\mu}^{(\lambda)}, \ \ 
\beta_{z,A+1}^{(\lambda)} 
 = \frac{1}{A} \sum_{\mu=1}^{A} \beta_{z,\mu}^{(\lambda)}\ \ \ {\rm for}\ \ z \in \{0,1\}, \nonumber \\ 
\alpha_{A+1}^{(n)} &=& \frac{1}{A} \sum_{\mu=1}^{A} \alpha_{\mu}^{(n)}.
\label{newhyperpara}
\end{eqnarray}
In addition, if some already-existing ensembles become empty (=no neuron) after update, we delete these ensembles and their hyperparameters. 

Now we consider the case that the membership label $\bm t^0 = \{t_1^0, t_2^0, \ldots, t_N^0 \}$ may be updated to new one $\bm t^* = \{ t_1^*, t_2^*, \ldots, t_N^* \}$ in MCMC. In this update, the ratio of conditional probabilities between $\bm t^0$ and $\bm t^*$ is calculated from equation (\ref{P(t, omega, s)}), which is necessary for acceptance rule of MCMC update,
\begin{eqnarray}
\frac{P(\bm t^*, \bm \omega, \bm s)}
     {P(\bm t^0, \bm \omega, \bm s)}
 & = & \frac{\prod_{\mu=1}^{A}
       \left\lbrace
       {\Gamma}(G_{\mu}+\alpha_{\mu})\prod_{z = \{0, 1\}}
       \it{B}({T}_{\mu}^{z1},{T}_{\mu}^{z0})
       \right\rbrace
       \vert_{\bm t = \bm t^*}}
       {\prod_{\mu=1}^{A}
       \left\lbrace
       {\Gamma}(G_{\mu}+\alpha_{\mu})\prod_{z = \{0, 1\}}
       \it{B}({T}_{\mu}^{z1},{T}_{\mu}^{z0})
       \right\rbrace
       \vert_{\bm t = \bm t^0}} 
       \nonumber \\ 
 & &   \cdot 
       \left[
       \frac{{\Gamma}(\sum_{\mu=1}^{A+1}\alpha_{\mu}^{(n)})}
            {{\Gamma}(\sum_{\mu=1}^{A}\alpha_{\mu}^{(n)})}
       \cdot     
       \frac{{\Gamma}(G_{A+1}+\alpha_{A+1}^{(n)})} 
            {{\Gamma}(\alpha_{A+1}^{(n)})}
       \right. \nonumber \\
 & &   \left. \cdot 
       \frac{B(H_{A+1},\bar{H}_{{A+1}})
             \cdot\prod_{z = \{0, 1\}}B(T_{A+1}^{z1},T_{A+1}^{z0})}
            {B(\alpha_{A+1}^{(p)},\beta_{A+1}^{(p)})
             \cdot\prod_{z = \{0, 1\}}
             B(\alpha_{z,A+1}^{(\lambda)},\beta_{z,A+1}^{(\lambda)})}   
       \right].
\label{transient rate}
\end{eqnarray}
The factor in the square bracket is the contribution from transient neurons to the new $(A+1)$th ensemble. If there is no neuron to the new ensemble, the factor in the square bracket vanishes because the denominator and the numerator cancel out. 

For Metropolis-Hastings update rule, we also need to define proposal distribution from the $t_{i}^{0}$th ensemble to the $t_{i}^{*}$th, $Q_{i}(t_{i}^{\ast} | t_{i}^{0})$, and its reverse process $Q_i(t_{i}^{0} | t_{i}^{\ast})$ for the $i$th neuron. These probabilities are calculated
to satisfy detailed balance condition as 
\begin{eqnarray}
Q_{i}(t_{i}^{\ast} | t_{i}^{0}) & = & \begin{cases}
                                     \displaystyle \frac{G_{t_{i}^*}^{(\backslash i)}}{q_{\alpha}^{[\gamma]}+ N - 1}
                                     & {\rm for} \ \ t_{i}^* = 1, \ldots, A, \vspace{1mm} \\
                                     \displaystyle \frac{q_{\alpha}^{[\gamma]}}{q_{\alpha}^{[\gamma]}+ N - 1}
                                     & {\rm for} \ \ t_{i}^* = A + 1,    
                                     \end{cases} \label{proposal} \\
Q_{i}(t_{i}^{0} | t_{i}^*) & = & \frac{G_{t_{i}^{0}}^{(\backslash i)}}{N-1},
\end{eqnarray}
for the $\gamma$th MCMC stage, where equation (\ref{proposal}) is the same as (\ref{DP}). If multiple neurons move simultaneously, we must consider the product of the probabilities above for all transient neurons. As a result, acceptance rate from the membership label $\bm t^0$ to $\bm t^*$ is written as 
\begin{eqnarray}
a(\bm t^*, \bm t_0) &=& {\rm min}
				      \left\lbrace	
                      1,\ 
                      \frac{P(\bm t^{\ast}, \bm \omega, \bm s)}
                           {P(\bm t^{0}, \bm \omega, \bm s)}  
					  \frac{Q(\bm t^{0} | \bm t^{\ast})}
						   {Q(\bm t^{\ast} | \bm t^{0})}
                      \right\rbrace, \\
\label{acceptance rate}                      
{\rm where} \ \ \ 
Q (\bm t^0 | \bm t^*) &=& \prod_{i=1}^N Q_i (t_i^0 | t_i^*), \ \ 
Q (\bm t^* | \bm t^0) = \prod_{i=1}^N Q_i (t_i^* | t_i^0).
\end{eqnarray}

Finally, we update hyperparameters of $\{ \bm p, \bm \lambda, \bm n \}$ for remaining ensembles. For hyperparameter update, we introduce learning rate $\varepsilon^{[\gamma]}$, where $\gamma$ is the stage of MCMC update as in (\ref{qdecay}), to control the effect of simulated annealing. Here we use sigmoid function for $\varepsilon^{[\gamma]}$ because it is bounded and smooth,
\begin{equation}
\varepsilon^{[\gamma]} = \frac{1}{1 + e^{- \frac{\gamma}{\tau}}}.
\end{equation}
The decay constant $\tau$ is the same as in (\ref{qdecay}). Note that we do not need to introduce additional hyperparameter for the learning rate. Following the update rule in the original\cite{Giovanni:Thomas:Martin}, hyperparameters should be updated with learning rate $\varepsilon^{[\gamma]}$ as 
\begin{eqnarray}
\tilde{\alpha}_{\mu}^{(p)} & = & 
\alpha_{\mu}^{(p)} + \varepsilon^{[\gamma]}
                     \left(
                     \sum_{k=1}^{M} \omega_{k \mu}
                     \right),\ \
\tilde{\beta}_{\mu}^{(p)} = 
\beta_{\mu}^{(p)} + \varepsilon^{[\gamma]}
                    \left(
                    \sum_{k=1}^{M} (1 - \omega_{k \mu})
                    \right), \nonumber \\
\tilde{\alpha}_{z, \mu}^{(\lambda)} & = & 
\alpha_{z, \mu}^{(\lambda)} + \varepsilon^{[\gamma]} 
                              \left(
                              \sum_{k=1}^{M}
                              \left(
                              \sum_{i \in \bm \mu}
                              \delta_{z, \omega_{k \mu}}
                              \delta_{1, s_{i k}}
                              \right)
                              \right), \nonumber \\
\tilde{\beta}_{z, \mu}^{(\lambda)} & = & 
\beta_{z, \mu}^{(\lambda)} + \varepsilon^{[\gamma]} 
                             \left(
                             \sum_{k=1}^{M}
                             \left(
                             \sum_{i \in \bm \mu}
                             \delta_{z, \omega_{k \mu}}
                             \delta_{0, s_{i k}}
                             \right)
                             \right), \nonumber \\
\tilde{\alpha}_{\mu}^{(n)} & = & 
\alpha_{\mu}^{(n)} + \varepsilon^{[\gamma]} G_{\mu},
\label{hyperpara-update}
\end{eqnarray}
where tilde means updated hyperparameter. By introducing learning rate, hyperparameter update is suppressed at early stage of MCMC, when the number of ensembles $A$ is frequently changed instead.

To conclude, we summarize our MCMC process as the pseudo code in Algorithm \ref{alg.update}. The symbol $\bm \omega_{\backslash k\mu}$ means the set of parameters $\bm \omega$ excepting $\omega_{k \mu}$ for describing collapsed Gibbs sampling.

\begin{algorithm}[t]                       
\caption{Inference of ensembles activity and the number of ensembles}         
\label{alg.update}                          
\begin{algorithmic}                  
\STATE{initialize $\bm \omega$ and $\bm t$ 
}
\WHILE{the number of ensembles $A$ converges}
\FOR{each ensemble $\mu \in [1, A]$, $k \in [1, M] $}
\STATE{draw $\omega_{k \mu} \sim 
             P(\omega_{k \mu}=1|t, \bm \omega_{\backslash k \mu}, s)$}
\ENDFOR
\FOR{each neuron $i \in [1, N]$}
\STATE{draw destination ensemble $t_i^{\ast} \sim q(t_i^{\ast})$ in equation (\ref{DP})}
\IF{$t_i^{\ast} = A+1$}
\STATE{$A \rightarrow A+1$}
\STATE{give new hyperparameters as equation (\ref{newhyperpara})}
\ENDIF
\ENDFOR
\FOR{each neuron $i \in [1, N]$}
\STATE{draw $\bm t \sim a(\bm t^{\ast}, \bm t^{0})$ 
       in equation (\ref{acceptance rate})}
\ENDFOR
\FOR{each ensemble $\mu \in [1, A]$}
\IF{$G_{\mu} = 0$}
\STATE{delete the $\mu$th ensemble and its hyperparameters}
\ENDIF
\ENDFOR
\STATE{update hyperparameters as equation (\ref{hyperpara-update})}
\ENDWHILE
\end{algorithmic}
\end{algorithm}

\section{Experiment}
\subsection{Generative model for synthetic data}
Before discussion on utility of our algorithm, we summarize how to generate synthetic neuronal activity data for our numerical experiment. In our generative model, neuronal activities are closely related to ensemble activities as in equation (\ref{condactivity}), and the relation between neuronal/ensemble activities is characterized by conditional activity rate $\bm \lambda$. Hence, to generate synthetic data, we first divide all neurons into ensembles. Next we generate ensemble activity data $\bm \omega$ based on ensemble activity parameter $\bm p$. In the last we determine activity of each neuron $\bm s$ by conditional activity rate $\bm \lambda$.

The algorithm of synthetic data generation is summarized as the pseudo code  in Algorithm \ref{alg.generate}. Detail can be found in the original work as well\cite{Giovanni:Thomas:Martin}.

\begin{algorithm}                      
\caption{Generation of synthetic neuronal activity data}         
\label{alg.generate}                    
\begin{algorithmic}                  
\STATE{set all $\bm \omega$ and $\bm s$ to be 0}
\FOR{each ensemble $\mu \in [1, A]$, $k \in [1, M] $}
\STATE{draw $\omega_{k\mu} \sim P(\omega_{k\mu})$}
\ENDFOR
\FOR{each neuron $i \in [1, N]$}
\STATE{deal the $i$th neuron to an ensemble}
\ENDFOR
\FOR{each neuron $i \in [1, N]$, $k \in [1, M]$}
\IF{$\omega_{kt_{i}} = 1$}
\STATE{draw $s_{ik} \sim P(s_{ik}|\omega_{kt_{i}}=1)$}
\ELSE
\STATE{draw $s_{ik} \sim P(s_{ik}|\omega_{kt_{i}}=0)$}
\ENDIF
\ENDFOR
\end{algorithmic}
\end{algorithm}

\begin{table}
\caption{The condition of synthetic data generation}
\begin{center}
\begin{tabular}{c|c} \hline 
parameter & value \\ \hline
the number of neurons & $N=100$ \\
the number of ensembles & $A=10$ \\
ensemble activity rate & $p_{\mu}=0.1\ \ (\forall \mu)$ \\
conditional activity rate \ & \
$\lambda_{\mu}(0)=0.01$,  $\lambda_{\mu}(1)=0.6$ \ $(\forall \mu)$\\ \hline 
\end{tabular}
\end{center}
\label{Table.generation condition}
\end{table}

\subsection{Numerical validation}
\label{Sec:NV}
\begin{figure}
\begin{picture}(0,520)
	\put(40,0){\includegraphics[scale=0.96]{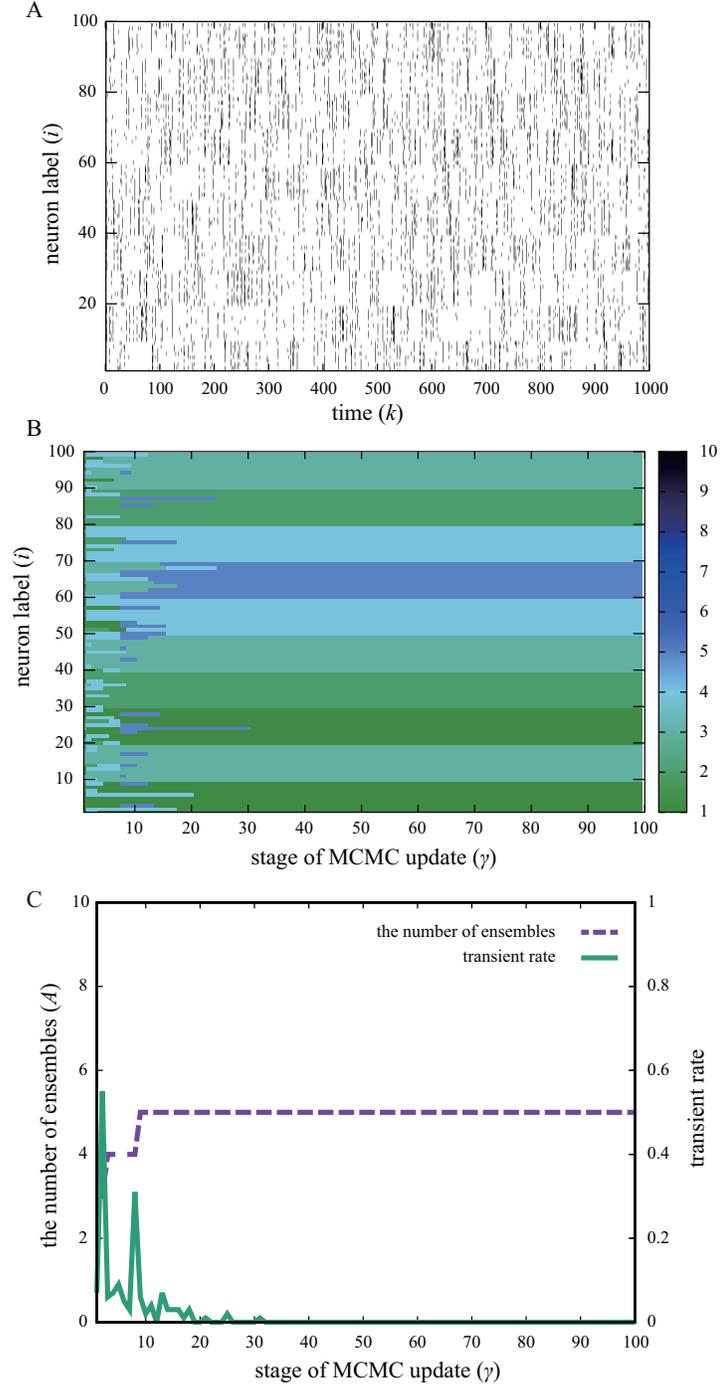}}
\end{picture}
\caption{Dynamical behavior of Algorithm \ref{alg.update} in MCMC:
         (A) Synthetic data of neuronal activity with $100$ neurons, $10$ ensembles and $1000$ time steps.
         black=active, white=inactive.
	     (B) Behavior of ensemble membership label.  
	     (C) Behavior of the number of ensembles (broken) and transient rate (solid).}
\label{Fig.result}
\end{figure}

To validate our method, we conduct numerical experiment for neuronal ensemble inference. In the experiment, we first generate synthetic data by Algorithm \ref{alg.generate}, then we extract the information of ensembles by Algorithm \ref{alg.update}. We illustrate a sample of activity matrix $\bm s$ in Fig. \ref{Fig.result}A, which is obtained by Algorithm \ref{alg.generate}. In this sample we have 10 ensembles and 100 neurons, where each ensemble has equally 10 neurons. The vertical axis shows the neuron label, which is sorted by neuronal membership label $\bm t$. We can easily see the structure of 10 ensembles, however we should note that it cannot be seen if neuron labels are randomly permuted. The parameters for synthetic data generation are summarized in Table \ref{Table.generation condition}. All ensembles/neurons are generated with the same ensemble activity rate $\bm p$ and conditional activity rate $\bm \lambda$. 

For ensemble inference, we use the data in Fig. \ref{Fig.result}A as input activity $\bm s$. As shown in Algorithm \ref{alg.update}, we repeatedly update ensemble activity $\bm \omega$, ensemble membership label $\bm t$, and hyperparameters till the number of transient neurons in DP becomes sufficiently small. In this experiment, we set initial number of ensembles $A=3$, decay constant $\tau=10$, initial transient parameter $q_{\alpha}^{[0]}=100$ and hyperparameters $\alpha_{\mu}^{(p)}=100, \beta_{\mu}^{(p)}=100, \alpha_{z,\mu}^{(\lambda)}=100, \beta_{z,\mu}^{(\lambda)}=100, \alpha_{\mu}^{(n)}=100$ for all $\mu,z$. At initialization step, we randomly assign initial membership label to each neuron uniformly between 1 to $A$.

In Fig. \ref{Fig.result}B, we show a typical example of dynamical membership label behavior in MCMC by the heat map, where the horizontal axis indicates the stage of MCMC update. The colors in the heat map distinguish ensemble numbers \{$1, 2, \ldots, A$\}. The original ensemble structure is clearly obtained at most after 40 MCMC stages.  
 
In Fig. \ref{Fig.result}C, we show the behavior of the number of ensembles $A$ and transient rate in MCMC, which is calculated from the result in Fig. \ref{Fig.result}B.  Transient rate is defined by 
\begin{equation}
{\rm transient\ rate} = \frac{1}{N} \sum_{i=1}^N
\left( 1 - \delta_{t_i^{[\gamma]}, t_i^{[\gamma-1]} } \right)
\end{equation}
at the $\gamma$th MCMC stage, namely the fraction of transient neurons between the $(\gamma-1)$th and the $\gamma$th MCMC stages. The broken line represents the number of ensembles and the solid line represents transient rate. From Fig. \ref{Fig.result}C, we find that the number of ensembles and ensemble membership label converge after 40 MCMC stages.

We should note that the number of final ensembles is 5, which is smaller than the ground-truth value 10. This is because some ensembles are merged in the final result. To manage this problem, we use our algorithm repeatedly with another initial membership label, which leads to different ensemble structure. Even in this case, we will obtain blockwise ensemble structure again like in Fig. \ref{Fig.result}B, while the number of ensemble is still smaller than 10. However, we should note that {\it other} ensembles are merged under different initial membership label. Therefore, we can obtain the original 10-ensemble structure exactly by combining several MCMC results with different initial membership label.

When we change conditional activity rate $\bm \lambda$, which controls coherence or noise, the ensemble inference becomes easy/hard. Even under hard condition or noisy case, we can still obtain nearly correct ensemble structure, and noise can be removed as much as possible. In addition, even if the sizes of ground-truth ensembles are not equal unlike Fig. \ref{Fig.result}A, we can infer correct ensemble structure.

\subsection{Comparison with the original algorithm}
We compare our algorithm and the original one in Ref.\cite{Giovanni:Thomas:Martin}. For comparison, we use the same synthetic data in Fig. \ref{Fig.result}A. In the original algorithm, we do not use simultaneous update rule for multiple neurons to new ensemble, nor simulated annealing idea for hyperparameters (or take the limit of $\tau \rightarrow + \infty$).

\begin{figure}
\begin{picture}(0,360)
	\put(40,0){\includegraphics[scale=0.96]{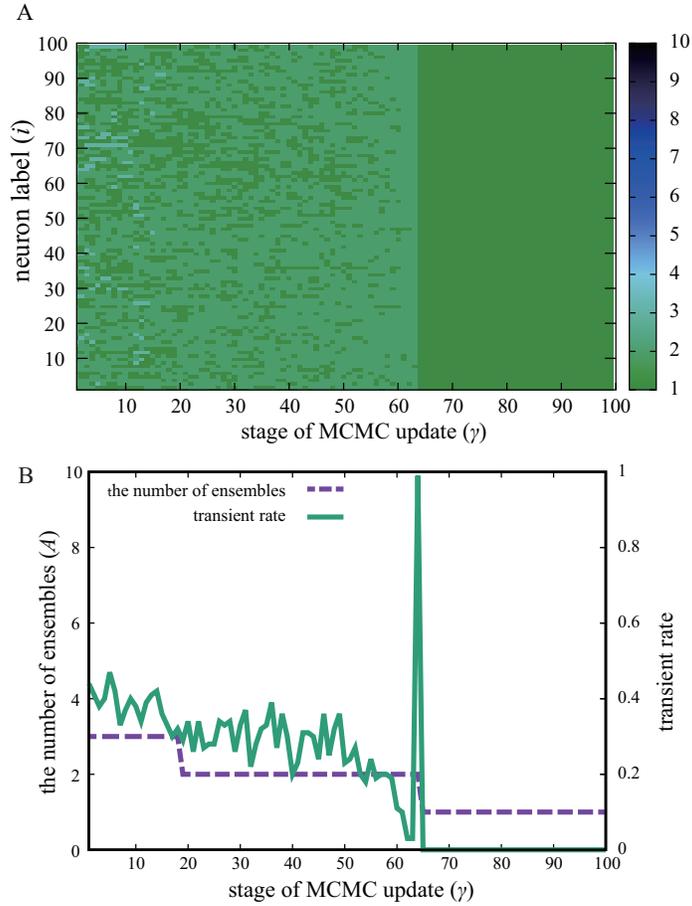}}
\end{picture}
\caption{Dynamical behavior of the original algorithm in Ref.\cite{Giovanni:Thomas:Martin} in MCMC:
(A) Behavior of ensemble membership label.
(B) Behavior of the number of ensembles (broken) and transient rate (solid).}
\label{Fig.result2}
\end{figure}

A typical example of dynamics is shown in Fig. \ref{Fig.result2}. In Fig. \ref{Fig.result2}A, we start with small number of ensembles $(A=3)$ like our algorithm in Sect. \ref{Sec:NV}. However, we cannot obtain blockwise structure. All ensembles are merged into one at late MCMC stage. In Fig. \ref{Fig.result2}B, we show the number of ensembles and transient rate in the result of Fig. \ref{Fig.result2}A, which exhibits slow convergence of transient rate. We verify that this result does not depend on the initial condition such as ensemble membership label. From these results, we conclude that the original algorithm will get stuck in bad local maximum of Bayesian inference, when we start with small initial number of ensembles.

\newpage
\section{Discussion and perspective}
In this work, we proposed Bayesian inference algorithm with faster convergence. In our method, we introduced the update rule to new ensemble for multiple neurons and the idea of simulated annealing, to avoid bad local maximum solution of Bayesian inference. For simulated annealing, we introduced decay constant $\tau$ to control annealing schedules of transient probability and learning rate. As a consequence, we find that our method can successfully obtain blockwise neuronal ensemble structure by numerical experiment for synthetic data, even with small initial number of ensembles. We also compare our algorithm with the original one, and the result indicates our algorithm has advantage for finding correct ensemble solution.

Note that our method in this work focuses on neuronal ensemble identification, not for the detail of neural network structure like connection. However, we believe that our idea for ensemble identification will be helpful for understanding whole structure of biological neural network. Moreover, by further improvement of our method, we think that we can construct Bayesian inference framework for the detail of neural network structure.

Several issues are remained as future works. First, the additional hyperparameter $\tau$ may also be useful for finding hierarchical ensemble structure. If $\tau$ is set to be large or annealing schedule is slow, we can obtain finer ensemble structure, while it requires many MCMC stages till convergence. On the other hand, if $\tau$ is small we can obtain ensemble structure more faster. However, only large scale structure will be found and fine structure will be neglected. Even in this case, we can obtain fine structure if we use this method repeatedly, as mentioned in Sect. \ref{Sec:NV}. We should investigate the role of the hyperparameter $\tau$ in more detail.

Second, we should apply our algorithm to real neuronal activity data, which we are planning at present. One of the problems for application is that real experimental data of neuronal activity is often continuous, not binary like our formulation. The natural idea for application to continuous data is to binarize real activity data, however this may neglect significant information in neuronal activity. Another idea is to generalize our formalism to continuous activity data, and for this idea we must modify generative model for continuous data. We should consider which strategy is more appropriate for application to real activity data.
\section{Acknowledgements}
We appreciate the comments from Giovanni Diana and Yuishi Iwasaki.
This work is supported by KAKENHI Nos. 18K11175, 19K12178.

\end{document}